\documentclass[10pt,a4paper]{article}
\usepackage{times}
\usepackage{graphics}
\usepackage{graphicx}
\usepackage{float}
\restylefloat{figure}

\begin{document}
\large
\date{ }

\begin{center}

{\Large A Differential Time-of-flight Spectrometer of Very Slow Neutrons}

\vskip 1.cm

Yu. N. Pokotilovski\footnote{Corresponding author; e-mail: pokot@nf.jinr.ru}
\vskip 0.1cm
            Joint Institute for Nuclear Research\\
              141980 Dubna, Moscow region, Russia\\
\vskip 0.4cm
M. I. Novopoltsev
\vskip 0.1cm
         Mordovian State University\\
        430000 Saransk, Russia
\vskip 0.4cm
P. Geltenbort and Th. Brenner
\vskip 0.1cm
         Institut Laue-Langevin,\\
    BP 156, 38042 Cedex 9  Grenoble, France

\vspace{5mm}

{\bf Abstract}

\begin{minipage}{120mm}

\vskip 0.5cm

 A time-of-flight spectrometer of neutrons in the energy range (0.05 -- 2.5)
$\mu$eV is described.
 This spectrometer has been tested my measuring the total and differential
neutron cross sections for a number of materials: Al, Cu, $^{6}$LiF, Si, Zr,
teflon, polyethylene and liquid fluoropolymers, that are essential for
experiments in the physics of ultracold neutrons.
\end{minipage}
\end{center}

\vskip 0.3cm


\vskip 0.2cm


\vskip 0.6cm

\section{Introduction}
 Ultracold neutrons (UCN) \cite{ucn} with energy less than $\sim$ 0.25 $\mu$eV
can be confined within closed volumes over periods of time as long as the
neutron $\beta-$decay.
 This phenomenon has found well known applications in the investigations of
fundamental properties of the neutron \cite{ILL}: searching for the neutron
electric dipole moment, measuring the neutron lifetime, investigation of
angular correlations in neutron decay.
 The examples of applications of very low energy neutrons in the $\mu$eV energy
range and lower for studies in condensed matter physics are rather scarce (see
for example the reviews \cite{Gol,Mich}).

 Owing to the very low neutron energy it is possible to reach rather high
energy and momentum resolution:
$\sim $1 neV and $\sim$ 10$^{-4}$ \AA$^{-1}$ respectively.

 This high resolution opens up the obvious new possibility in investigations of
supramolecular structure and dynamics of matter due to high sensitivity of very low
energy neutrons to scattering on the objects of the size 10-1000 \AA.
 The latter is particularly interesting foe studying in the soft matter, in
physics of polymer and biological science, where slow motion of large molecules
and clusters present one of most interesting and poorly investigated domains.

 The only reason for such a modest scale of application of very low energy
neutrons in condensed matter research is low intensity of neutron flux in
the very low energy range.

 The best up to now source of neutrons in this energy range - ILL
UCN turbine \cite{turb} yields $\sim$ 50 n/(cm$^{2}$s neV) in the maximum of
the neutron spectrum in vicinity 750 neV (v$\approx$12 m/s) and the UCN
density in the storage mode $\sim$40 cm$^{-3}$.

 Last years brought hope for significant progress in intensity of very low
energy neutron beams after commissioning of new UCN sources \cite{LA,FRM,PSI}.
 It is connected with possible use of the most effective cold moderators:
solid deuterium \cite{D2} or solid deuterocarbons at low temperatures
$\sim$5-10 K for production of very cold neutrons.

\section{Description of the spectrometer}
  As applied to UCN, the necessity to have a neutron guide tube in the flight
path in order to prevent neutrons from falling in the gravitational field (on a 1-m-long flight path, a neutron
with velocity 5 m/s falls down in the gravitational field by 20 cm) is a significant drawback of time-of-flight
neutron spectrometers with a horizontal location of the flight path \cite{horspe}.

 Nonmirror neutron reflection from the surface, proceeding with a some
probability even in the best mirror neutron guide tubes, changes the measured
longitudinal component of the neutron velocity.
 Moreover, in relatively long neutron guides with a large number of neutron
reflections from the walls of the neutron guide tubes on the path from the
chopper to the detector, diffuse scattering of a portion of neutrons causes the
neutron guide tube to act as a storage chamber.
 This phenomenon is intensified by the fact that neutrons with lowest energies
are reflected from the surface of the detector entrance windows.
 All these facts result in hold-up of scattered neutrons and distortions in the
measured spectrum.
 With this in mind we used previously in the flight path of the time-of-flight spectrometers with a horizontal
neutron guide the shortest possible neutron guide tubes ($<$50 cm long), a special nonreflecting neutron
detectors, and the fastest admissible for the UCN beam modulation in order to attain an acceptable resolution of
2-3 neV \cite{hor-my}.

 In the described here spectrometer, the flight path has a vertical geometry,
and the large area detector is located in a horizontal plane, so that the neutron guide tube is needless.
 Neutrons are accelerated on the flight path between the sample and the detector in the gravitational field;
as a result, the probability that a neutron will be reflected from the surface of the detector's entrance window
becomes almost zero.

 By contrast to  the vertical time-of-flight spectrometers of very slow neutrons
with the flight path located ahead of the sample \cite{Stespe,Perspe}, the
flight path of our spectrometer is disposed past the sample.
 This allows us to measure angular distributions of scattered neutrons and,
hypothetically in future, measure doubly differential cross sections $d^{2}\sigma/d\Omega d\epsilon$, provided
that the neutron beam incident on the sample is monochromatized.
 The axial geometry with respect to the vector of the gravitational field
permits the use of a coordinate detector for meausuring, without distortions,
the angular distribution of scattering of the collimated neutron beam.

 The angular distribution of scattered neutrons must also be monitored in
measurements of the total cross sections, in which the elastic cross section
component of the total cross section should be separated from the inelastic
component: at very low energies, elastic scattering basically proceeds on
inhomogeneities and can be distinguished in angular distributions.

 The schematic diagram of the spectrometer is shown in Fig. 1.

\begin{figure}
\begin{center}
\resizebox{20cm}{23cm}{\includegraphics[width=\textwidth]{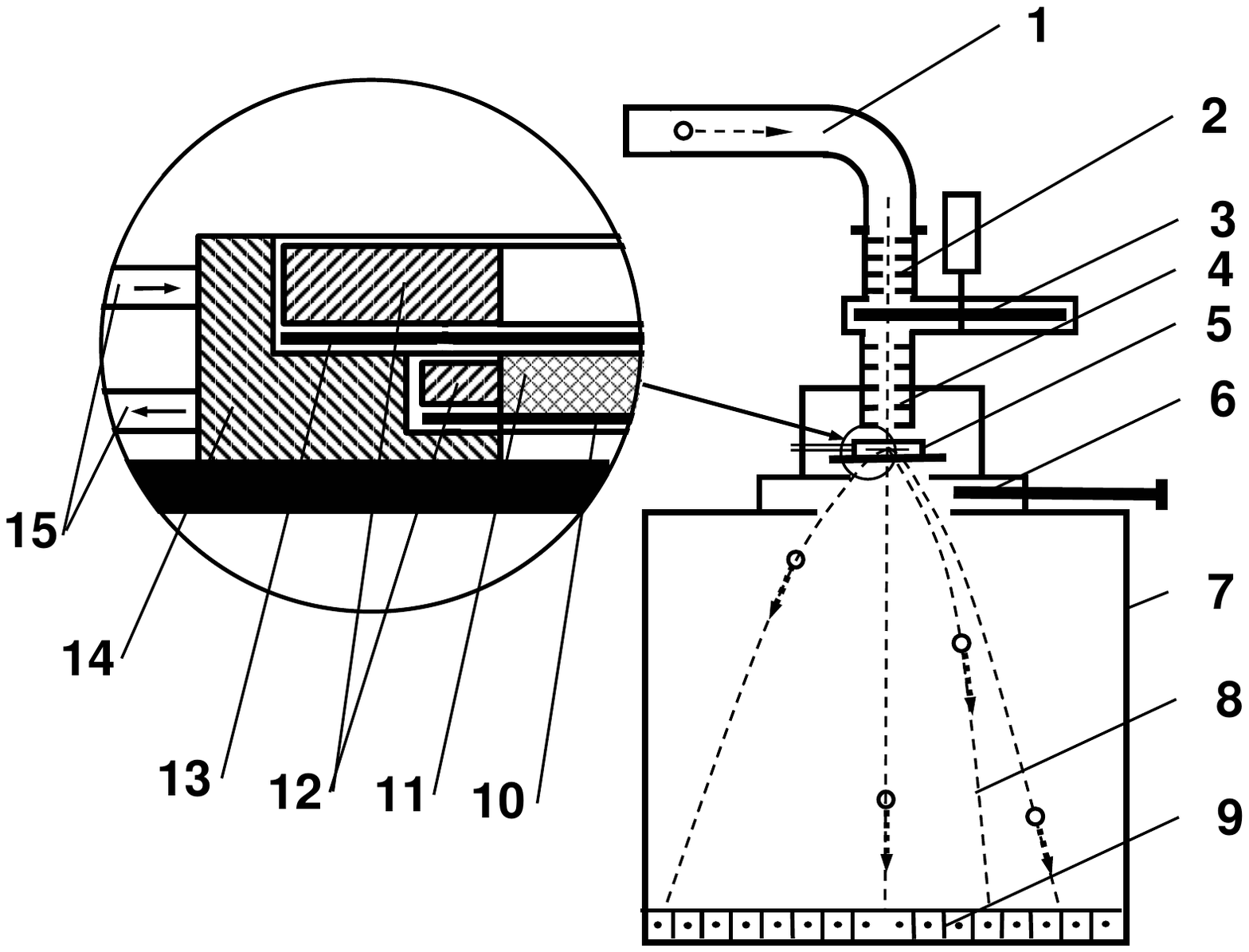}}
\end{center}
\caption{\Large The scheme of the spectrometer: 1 - neutron guide,
2 and 4 - collimators, 3 - chopper, 5 - sample holder, 6 - VAT valve,
7 - vacuum barrel, 8 - neutron trajectories, 9 - annular neutron position
sensitive detector, 10 and 13 - Si plates, 11 - sample, 12 - copper rings,
14 - copper holder, 15 - liquid nitrogen.}
\end{figure}

 Neutrons travel over neutron guide tube $1$, which is bent through 90$^o$,
toward collimator $2$ and then to chopper $3$ of the collimated beam.
 Additional collimation of the neutron beam is effected by collimator $4$
located past the chopper.
 Samples $11$ under investigation are located in sample holder $5$ inside the
vacuum chamber and can be cooled to low temperatures by pumping liquid
nitrogen through the sample holder.
 The sample holder parts shown in the inset in Fig. 1 are used for better
maintenance of the temperature conditions.
 Scattered neutrons that passed the sample enter vertical evacuated flight path
$7$ and move in parabolic trajectories $8$ in the Earth's gravitational field.
 The axis of these parabolic trajectories coincides with the vertical axis of
the ring detector.

 Neutrons are detected by gas-filled coordinate annular detector $9$, which
is located in a horizontal plane and has ring geometry (the radial width of the
ring counters is 3.4 cm, and maximum radius is 29 cm).
 The vertical distance from the chopper to the detector plane may be 79 cm or
160 cm, however the latter variant has not yet been tested.

 Between the vacuum time-of-flight barrel $7$ and the sample chamber $5$, there
is large-diameter vacuum gate valve $6$ (VAT DN 160) used to replace samples
without putting atmospheric air into the barrel.

 The replaceable chopper discs of 42.8 cm in diameter are made of 3 mm-thick
polyvinylchlorid coated with a gadolinium-containing paint.
 The chopper rotation frequency may varied from 3 to 8 Hz and is electronically
stabilized with precision $\sim 2\times 10^{-4}$.
 The chopper is driven by a SP-361 dc motor via a magnetic clutch.
 The trapezoidal windows in the chopper discs are 2, 4 and 8 cm wide, wich
corresponds to $\sim$ 2, 4 and 8\% of the open beam time with respect to the
chopper rotation period.

 A start signal from the chopper is produced by a LED via a slit drilled in the
disc on the side opposite to the chopper window.

 The 8-ring coordinate detector has a thickness of 1 cm and is filled with a
mixture of 50\% $^{3}$He and 50\% CF$_{4}$ ( we used $^{10}$BF$_{3}$ gas at the
first stage) to a pressure of $\sim$100 mbar.
 This corresponds to $n\sigma\approx$1.5 for neutrons with velocity 10 m/s.

 A 100-$\mu$m-thick aluminium foil supported by stainless steel grid is used
as a membrane of the entrance window of the coordinate detector.
 The inner surface of the time-of-flight volume and the surface of the
stainless steel grid are covered with polyethylene to prevent neutrons with very low energies from being
reflected.

 This spectrometer has been used to measure the total and differential neutron
cross sections.
 In the future, it is expected that doubly differential cross sections
$d^{2}\sigma$/d$\Omega d\epsilon$ will be measured.

 From the obvious relations for vertical and horizontal velocity components
\begin{equation}
v_{vert}=\frac{h}{t}-\frac{gt}{2}, \hspace{1cm} v_{hor}=\frac{r}{t},
\end{equation}
where $h$ is the vertical distance between the planes of the chopper and the
detector, $t$ is the neutron time-of-flight, $r$ is the radial coordinate of
the neutron detection point, and $g$ is the gravitational acceleration, it is
easy to obtain:
\begin{equation}
v=\Bigl(\frac{r^{2}+h^{2}}{t^{2}}-gh+(\frac{gt^{2}}{2})^{2}\Bigr)^{1/2}.
\end{equation}

 The uncertainties in velocities depending on uncertainties in $h$, $t$ and $r$,
are
\begin{equation}
\Delta v_{h}=\frac{2h-gt^{2}}{2vt^{2}}\Delta h,\quad
\Delta v_{t}=\frac{(gt^{2})^{2}-4(r^{2}+h^{2})}{4vt^{3}}\Delta t,\quad
\Delta v_{r}=\frac{r}{vt^{2}}\Delta r.
\end{equation}
 The error in determining the horizontal coordinate is determined by detector
ring width $\Delta R$ and the uncertainty due to angular spread $\Delta\theta$
of the incident neutron beam, which is equal to $\Delta\theta(r^{2}+h^{2})/h$.

 The uncertainties in the neutron energy are shown in Fig. 2 at $\Delta t$=3 ms
and 6 ms, $h$=79 cm, $\Delta h$=0.5 cm and $\Delta r=1.7$ cm.
 These uncertainties are weakly dependent on the radius of the detector R.
 The main contribution to these uncertainties comes from $\Delta t$.

 The expressions for the neutron wave vector transfer $Q$ with account of
gravity are:
\begin{equation}
sin\theta=\frac{r}{vt}, \quad \quad Q=2k\cdot sin(\theta/2),
\end{equation}
$k (\AA^{-1})=1.588\times 10^{-5} v (cm/s)$.
 The expressions for the uncertainties in Q, which include angle of divergence
of the incident beam $\Delta\theta$, and the uncertainties in $h$, $t$, and $r$
are rather sophisticated and are not shown here.
 The calculated energy dependence of Q and $\Delta Q$ are shown in Fig. 3 for
two values of radii of the neutron detector rings $R$ of 6.6 cm and 27 cm,
h=79 cm and $\Delta t$=3 ms.

\begin{figure}
\begin{center}
\includegraphics[width=\textwidth]{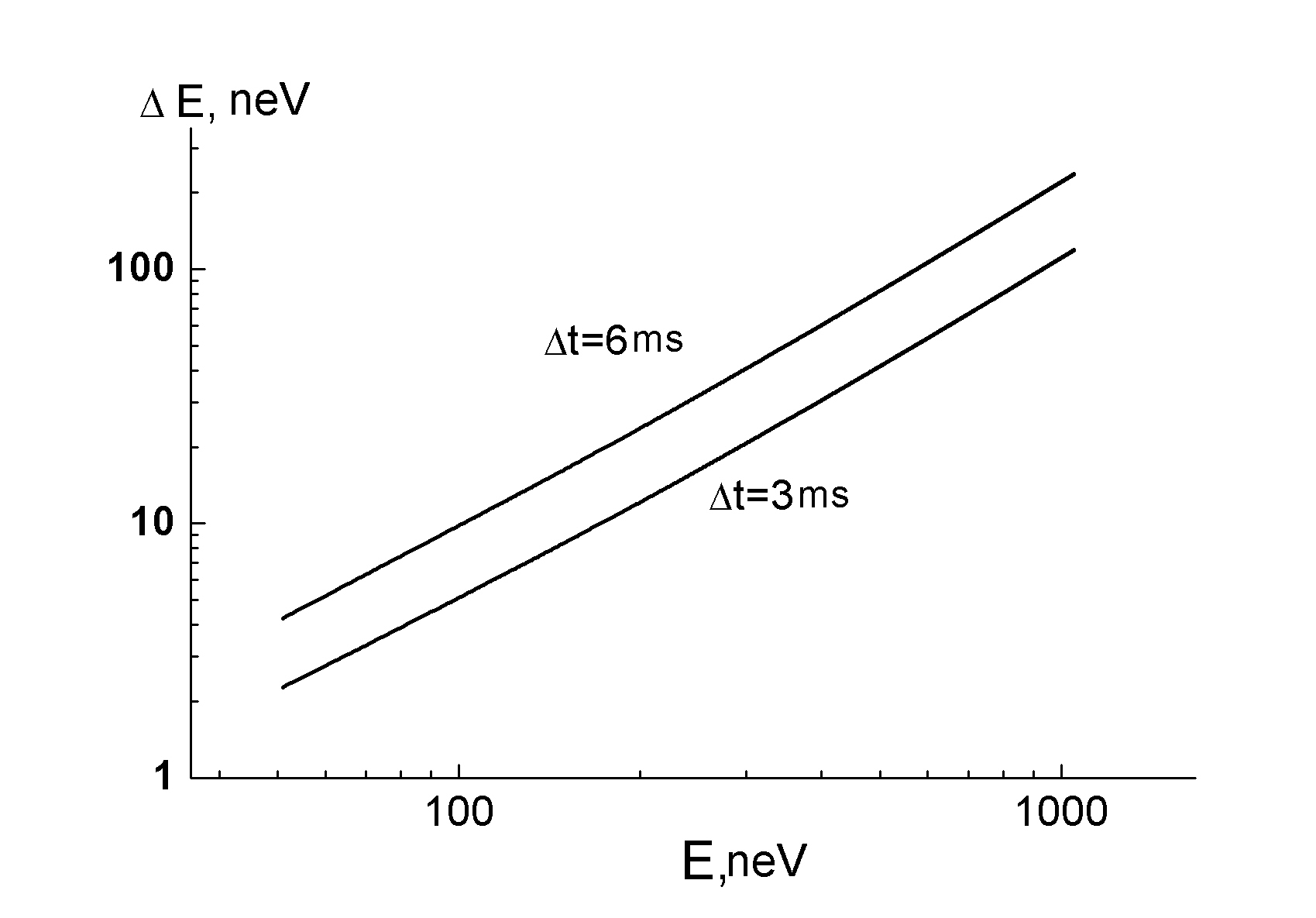}
\end{center}
\caption{\Large Calculated energy resolution of the spectrometer at a
flight path length of 79 cm.}
\end{figure}

\begin{figure}
\begin{center}
\includegraphics[width=\textwidth]{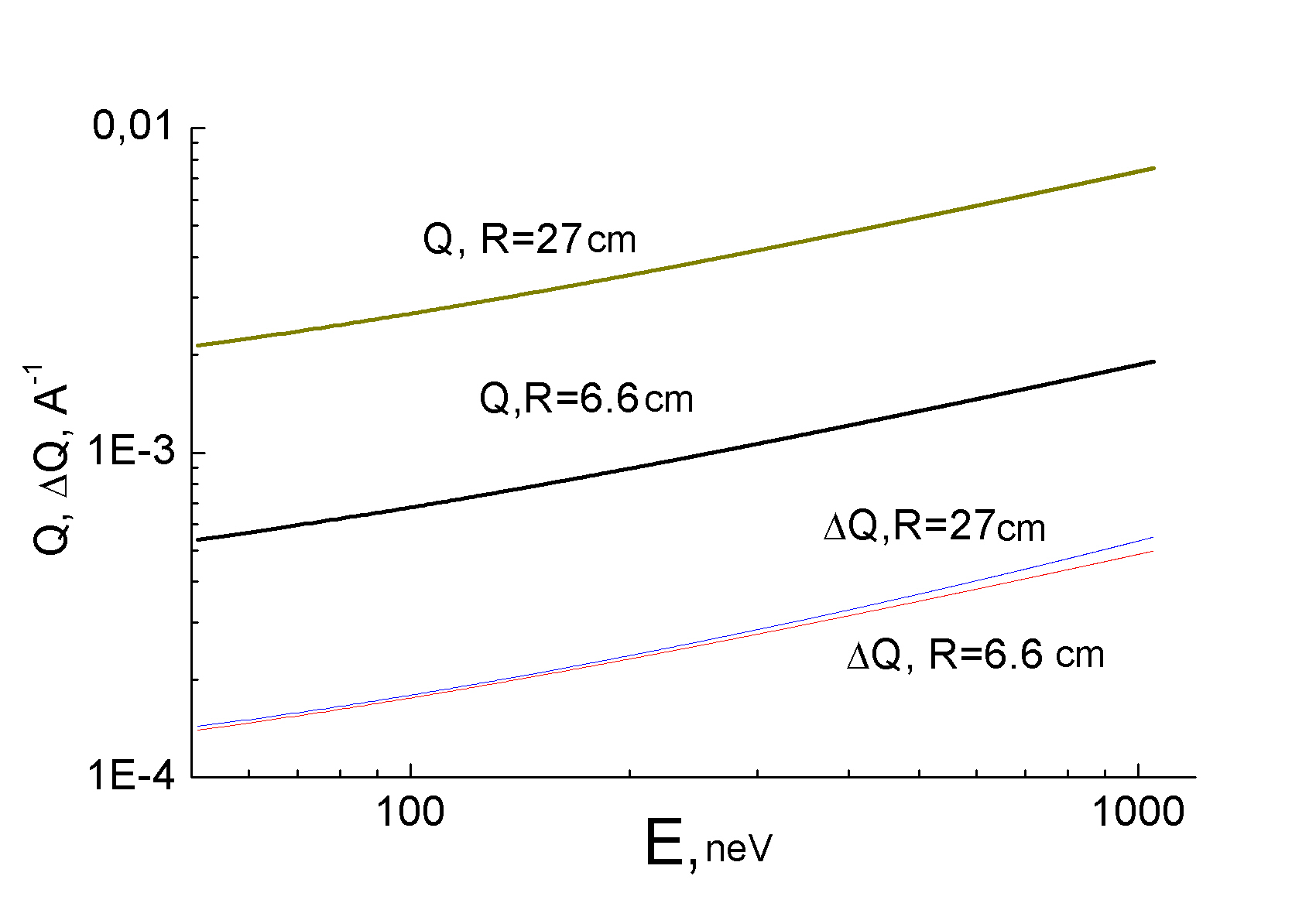}
\end{center}
\caption{\Large Calculated resolution in transmitted wave vector $Q$ (solid
lines) and $\Delta Q$ (dashed lines) at $\Delta t$=3 ms and a flight path length
of 79 cm.}
\end{figure}

 The tests included investigation of the properties of position sensitive gas
neutron detector filled with different gases: $^{10}$BF$_{3}$ and a mixture of
$^{3}$He and CF$_{4}$, optimization of the neutron collimators and test
measurements of the total and differential cross section for several samples.
 On-line monitoring and recording of pulse-hight spectra from all rings of the
detector was carried out in all measurements.
 Minimization of the background and optimization of the detector radiation
shielding under conditions of a low neutron flux were also essential.

 The total and differential cross sections were measured with rather good (for
this energy range) collimation of the incident beam, since, at collimation such
as this, the influence of the sample structure manifests itself to the maximum
degree.
 All the measured samples are relevant to the experiments in UCN physics and
therefore, a matter of some practical interest.
 The Al and Zr foils are used as entrance windows in UCN detectors, the exit
windows in the neutron sources, and, sometimes, as energy filters of neutrons
in some specific experiments with UCN.
 Polyethylene is used as an "ideal" UCN absorber, in particular, in gravitation
UCN spectrometers \cite{CH2,Rich} and for calibration of UCN losses in the UCN
storage chambers.
 Fluoropolymers are now thought to be the best coatings for walls of UCN
storage chambers \cite{PFPE}.
 $^{6}LiF$ layers are used in some UCN detectors \cite{LiF}.
 Being a very homogeneous material, silicon single-crystal wafers presented our
special interest and have found applications in our experiments with very cold
neutrons.
 Their good feature is that they do not broaden the collimated neutron beam and
are therefore suitable for sample chambers in neutron scattering experiments.

\section{Measurements and results}
 The measured scattering cross sections of very slow neutrons are shown in
Figs. 4-7.
 Figs. 4 and 5 illustrate the transmission of neutrons through an inhomogeneous
polycrystalline medium.

\begin{figure}
\begin{center}
\includegraphics[width=\textwidth]{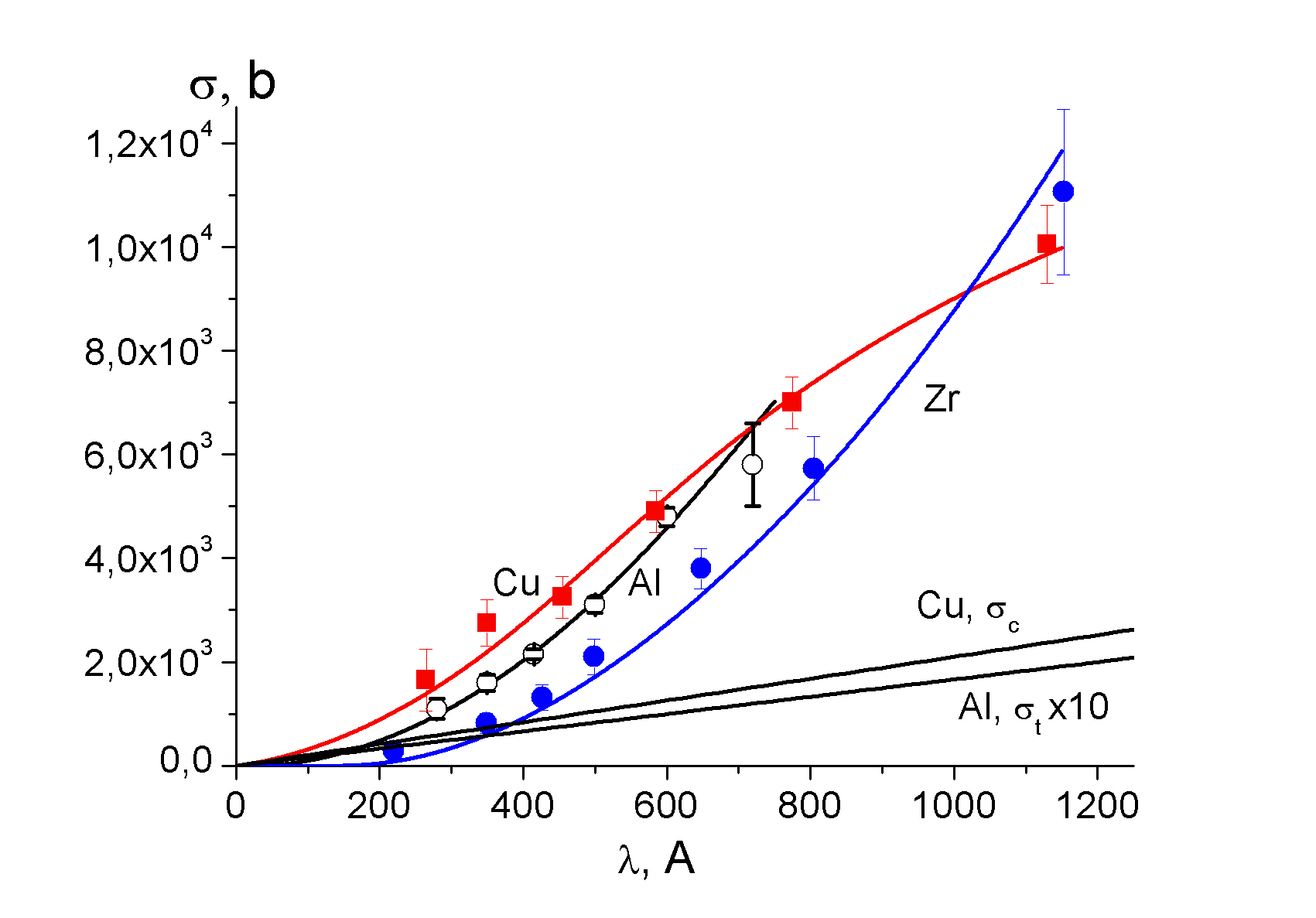}
\end{center}
\caption{\Large Total cross sections for Al (Goodfellow, 99.99\%, not annealed), Cu (Goodfellow, 99\%,
annealed), and Zr (Goodfellow, 99.8\%, 50 $\mu$m, not annealed) foils. The straight lines are the result of
extrapolation of cross section from the region of cold energies according to $1/v$ law for Al
($\sigma_{tot}$=1.8 b at an energy 0.6 mev \cite{Mug}, neutron capture cross section for Cu \cite{Mug} and cross
section for Zr ($\sigma_{tot}$=1.45 b at the energy 1 mev) \cite{Mug}. The curves passing through the
experimental points is result of fitting for Al according to Eq. (12) with $G_{0}=(6.5\pm 2.7) neV^{2},
\rho_{0}=(456\pm 190)$ \AA, for Cu according to Eq. (12) with $G_{0}=(13\pm 3) neV^{2}, \rho_{0}=(294\pm 35)$
\AA, and for Zr according to Eq. (9) with $G_{0}=(1.9\pm 0.12) neV^{2}, \rho_{0}=(840\pm 95)$ \AA.}
\end{figure}

 The transmission of slow neutrons through inhomogeneous media was investigated
earlier in \cite{Leng}.
 It was shown that, based on the transmission as a function of the neutron wave
length, one can deduce characteristic parameters of inhomogeneities, in
particular their size and density.
 Our measurements were performed at large wavelengths, up to
$\sim$10$^{3}$ \AA, for which the sensitivity to scattering on inhomogeneities
and clusters is substantially higher and the measured cross sections are
several orders of magnitude greater than the cross sections extrapolated from
the energy range of cold neutrons according to the $1/v$ law.

\begin{figure}
\begin{center}
\includegraphics[width=\textwidth]{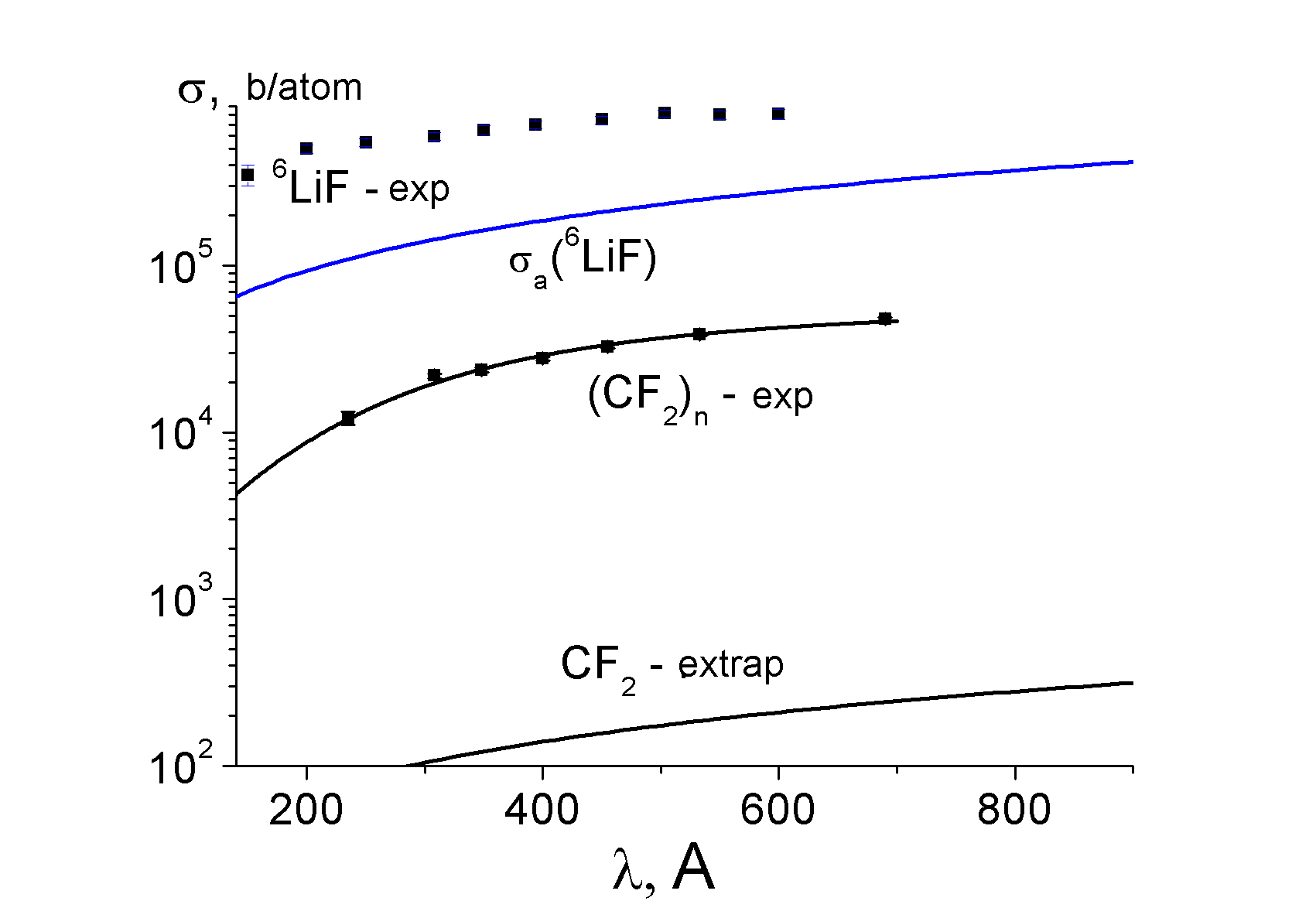}
\end{center}
\caption{\Large The total cross section for the Teflon film 10 $\mu$ m thick and $^{6}LiF$ layer 240 $\mu
g/cm^{2}$ thick. The solid lines are the results of the extrapolation of the cross section for Teflon from the
region of cold energies according to the $1/v$ law with $\sigma_{tot}$=5.18 b/atom for an energy 0.376 mev
\cite{Gran}, capture cross section for $^{6}LiF$, and the results of fitting of the experimental points for
Teflon according to Eq. (12) with $G_{0}=(400\pm 25) neV^{2}, \rho_{0}=(170\pm 15)$ \AA.}
\end{figure}

 It is known that in the Born approximation the differential macroscopic cross
section for elastic scattering of neutrons passing through an isotropic
inhomogeneous medium has the form \cite{inhom}:
\begin{equation}
\frac{d\Sigma_{el}}{d\Omega}=\frac{1}{\pi}\bigl(\frac{m}{\hbar^{2}}\bigr)^{2}
\int_{0}^{\infty}G(\rho)\frac{sin(q\rho)}{q\rho}\rho^{2}d\rho,
\end{equation}
where $m$ is the neutron mass, $q=|\vec k'-\vec k|$ is the change of the
neutron wave vector upon scattering, and
$G(\vec r,\vec r')=<\delta U(\vec r)\delta U(\vec r')>$, is the correlation
function of fluctuations of the local neutron-medium interaction potential, and
($\rho=|\vec r'-\vec r|$).

 This potential is
\begin{equation}
U=\frac{\hbar^{2}}{2m}\sum_{i} 4\pi N_{i}b_{i},
\end{equation}
where $N_{i}$ is the atomic density and $b_{i}$ is coherent scattering length of nuclei in the medium, so that
$\delta U(\vec r)=U(\vec r)-<U(\vec r)>$.

 After integrating over the solid angle with due account of the solid angle of
the neutron detector the total cross section is presented as:
\begin{equation}
\Sigma_{el}(k)=2\bigl(\frac{m}{\hbar^{2}}\bigr)^{2}\frac{1}{k^{2}}
\int_{0}^{\infty}G(\rho)[cos(2k\rho\cdot sin\theta_{0})-
cos(2k\rho)]d\rho,
\end{equation}
where 2$\theta_{0}$ -is the angle with which the neutron detector is viewed from the sample.

 For the exponential correlation function
\begin{equation}
G(\rho)=G(0)e^{-\rho/\rho_{0}},
\end{equation}
where $\rho_{0}$ is the correlation length, the expression for the total cross
section is
\begin{equation}
\Sigma_{el}(k)=2\bigl(\frac{m}{\hbar^{2}}\bigr)^{2}\frac{G_{0}\rho_{0}}{k^{2}}
\Bigl[\frac{1}{1+4k^{2}\rho_{0}^{2}sin^{2}\theta_{0}}-
\frac{1}{1+4k^{2}\rho_{0}^{2}}\Bigr],
\end{equation}
and the differential cross section has the form
\begin{equation}
\frac{d\Sigma_{el}}{dq}=\frac{2\pi q}{k^{2}}\frac{d\Sigma}{d\Omega}=
\frac{4q}{k^{2}}\bigl(\frac{m}{\hbar^{2}}\bigr)^{2}
\frac{G_{0}\rho_{0}^{3}}{(1+q^{2}\rho_{0}^{2})^{2}}.
\end{equation}

 For the Gaussian correlation function
\begin{equation}
G(\rho)=G(0)e^{-(\rho/\rho_{0})^{2}},
\end{equation}
the respective total cross section is
\begin{equation}
\Sigma_{el}(k)=\sqrt{\pi}\bigl(\frac{m}{\hbar^{2}}\bigr)^{2}
\frac{G_{0}\rho_{0}}{k^{2}}
\Bigl[e^{-(k\rho_{0}sin\theta_{0}/2)^{2}}-
e^{-(k\rho_{0}/2)^{2}}\Bigr]
\end{equation}
and the differential cross section is
\begin{equation}
\frac{d\Sigma_{el}}{dq}=\frac{2\pi q}{k^{2}}\frac{d\Sigma}{d\Omega}=
\frac{\sqrt{\pi}}{2}\bigl(\frac{m}{\hbar^{2}}\bigr)^{2}
\frac{G_{0}\rho_{0}^{3}q}{k^{2}}e^{-(\rho_{0}q/2)^{2}}.
\end{equation}

\begin{figure}
\begin{center}
\includegraphics[width=\textwidth]{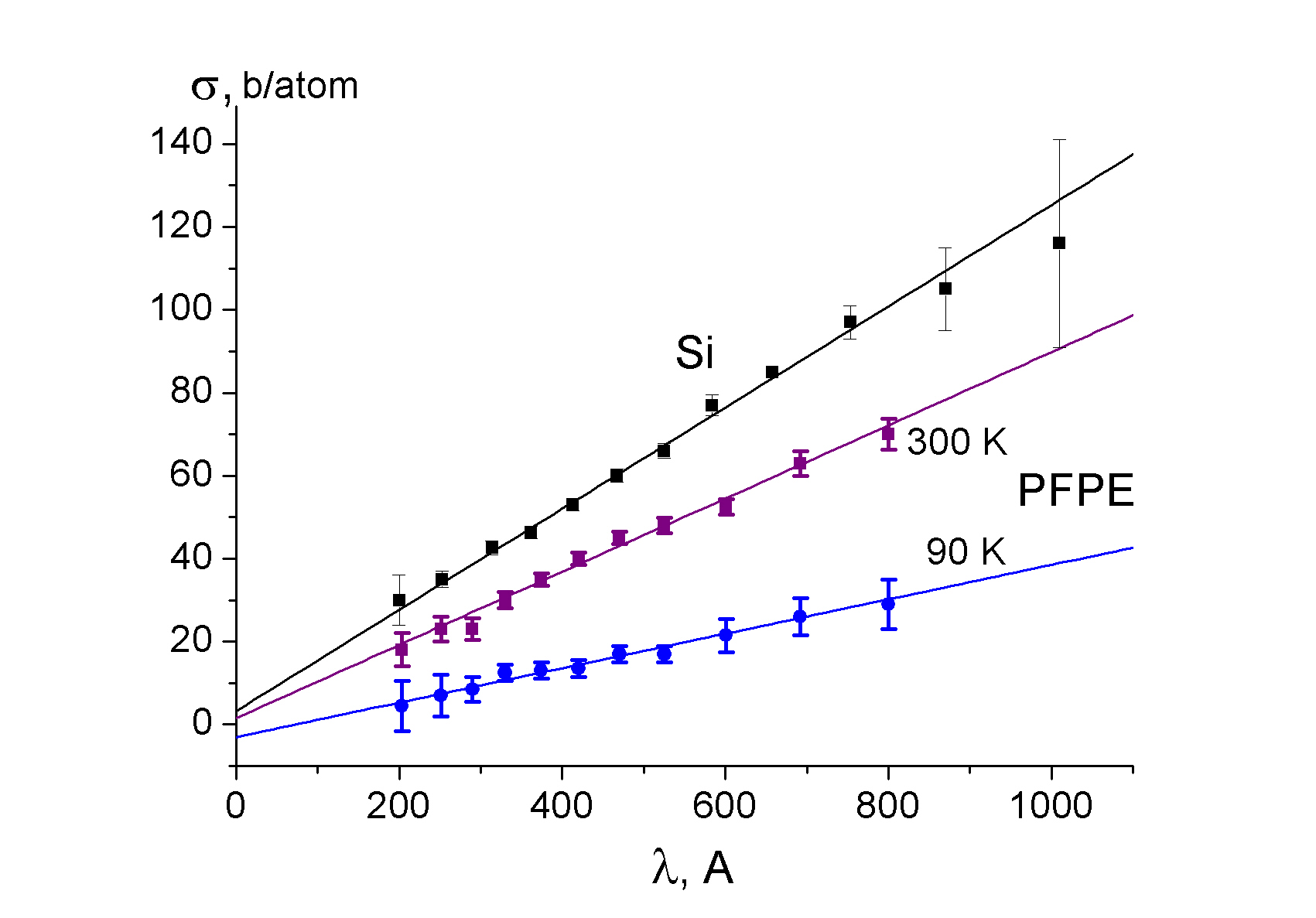}
\end{center}
\caption{\Large Total neutron cross sections for silicon and fluoropolymers.
The straight lines passing through the experimental points are the fits
according to Eq. (14)
for silicon with $\sigma_{0}=3.26\pm 2.3)$ b, $a=(0.122\pm 0.005)$ b/\AA,
for liquid fluoropolymer at room temperature with $\sigma_{0}=1.53\pm 2.2)$ b
and $a=(8.83\pm 0.5)\times 10^{-2}$ b/\AA,
and for the solid-state fluoropolymer at T=90 K $\sigma_{0}=3\pm 3)$ b and
$a=(4.15\pm 0.5)\times 10^{-2}$ b/\AA.}
\end{figure}

\begin{figure}
\begin{center}
\includegraphics[width=\textwidth]{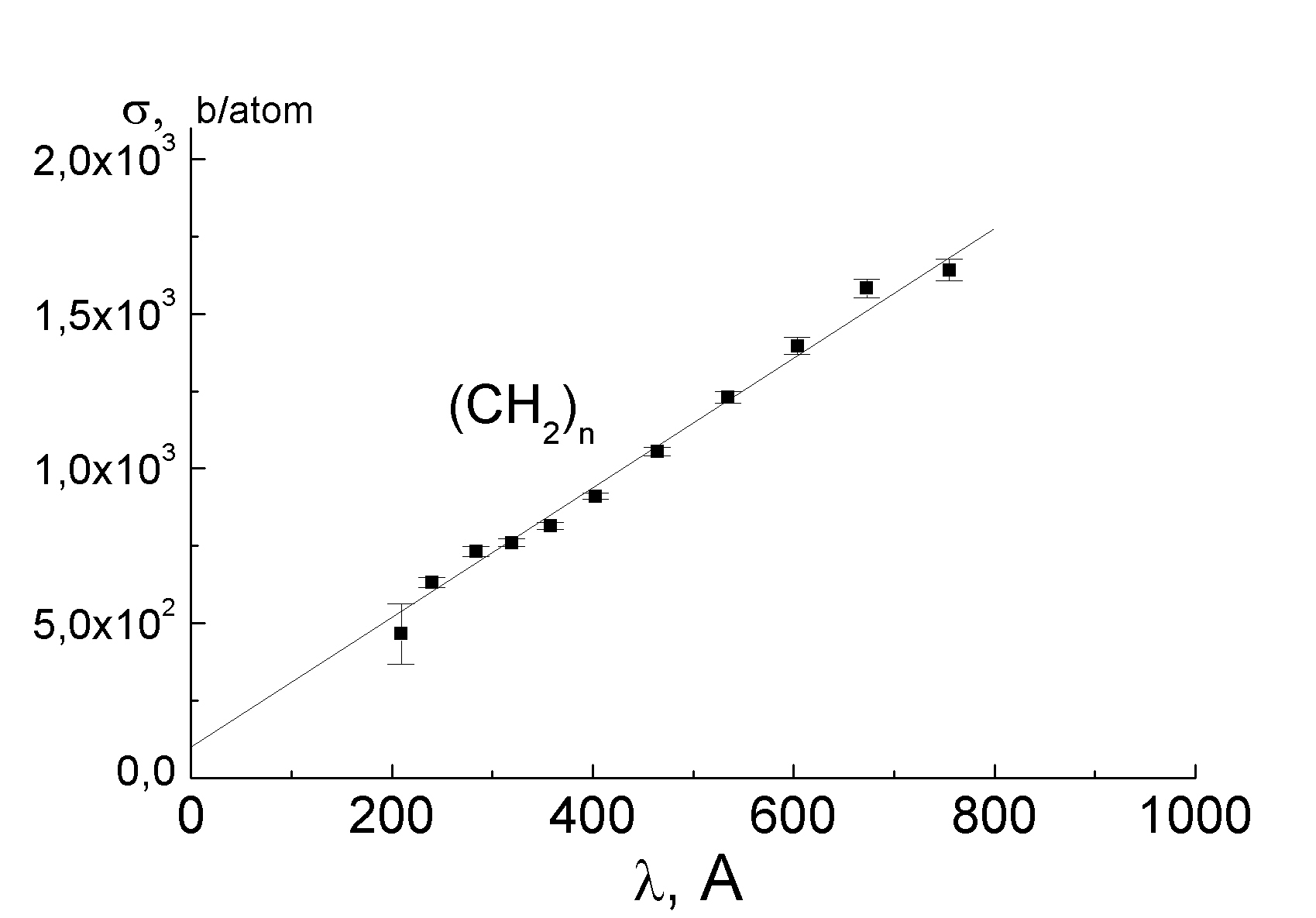}
\end{center}
\caption{\Large Total neutron cross section for the polyethylene film
7.88 mg/cm$^{2}$ thick.
The straight lines passing through the experimental points are the fits
according to Eq. (14) with $\sigma_{0}=96\pm 18)$ b, $a=(2.07\pm 0.04)$ b/\AA.}
\end{figure}

\begin{figure}
\begin{center}
\includegraphics[width=\textwidth]{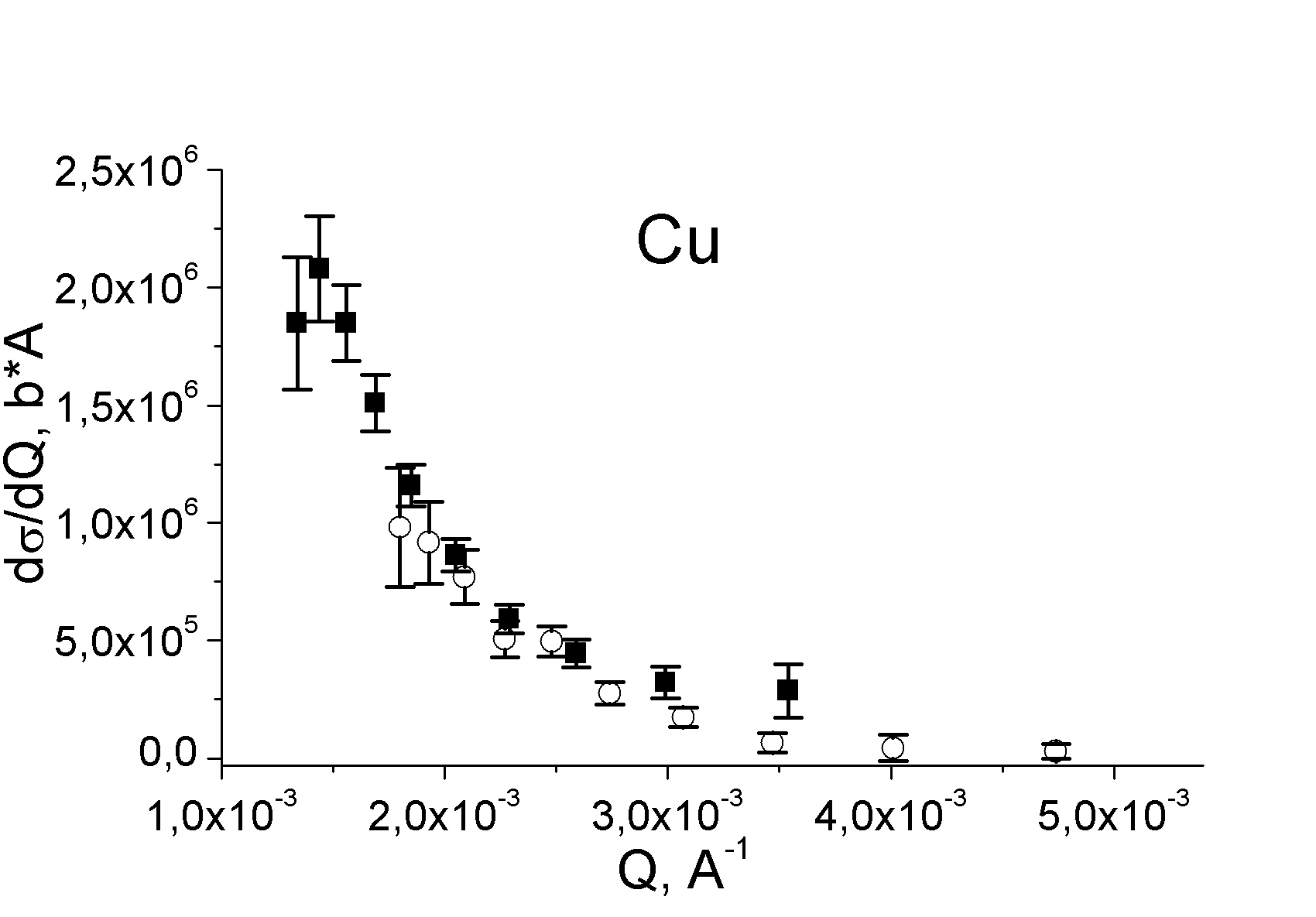}
\end{center}
\caption{\Large Differential cross section for copper.
The squares and the circles correspond to two different detector rings.}
\end{figure}

 These formulas were used to interpret measurements of the total and differential
cross sections for scattering from inhomogeneous samples (see also
\cite{Leng,inhom,Perfor}).
 It has been determined in our work that, in some cases, one of
the correlation functions (a Gaussian or exponential function) provides better
agreement with the experimental data, but, in other cases, the difference in
$\chi^{2}$ for these two models is inessential: both models represent
experimental data well.

 The following parameters of the correlation functions provide the best
description of the measured cross sections:
\begin{center}
Al (Gaussian): $G_{0}=(6.5\pm 2.7) neV^{2}, \rho_{0}=(456\pm 190)$\, \AA;

Cu (Gaussian): $G_{0}=(13\pm 3) neV^{2}, \rho_{0}=(294\pm 35)$\, \AA;

Zr (exponential): $G_{0}=(1.9\pm 0.12) neV^{2}, \rho_{0}=(840\pm 95)$\, \AA;

Teflon (Gaussian): $G_{0}=(400\pm 25) neV^{2}, \rho_{0}=(170\pm 15)$\, \AA.
\end{center}

 The total cross section of Teflon (Fig. 5) is two orders of magnitude larger than the
total cross section extrapolated from the region of cold neutrons \cite{Gran}
and demonstrates the very high degree of inhomogeneity of this material.

 Measuring the transmission of thin $^{6}LiF$ layers (with 90\% $^{6}Li$
content, we see that, even in this case, the total cross section is much
greater than the extrapolated very large cross section for neutron capture by
$^{6}Li$ nuclei (Fig. 5).
 This indicates that the $^{6}LiF$ layers obtained using the method of thermal
evaporation in vacuum have a very high degree of inhomogeneity.
 Such a strong scattering from inhomogeneities will result in significant albedo
of very slow neutrons and in a decrease in the UCN  detector efficiency
\cite{ LiF}, particularly for neutrons near the potential barrier of the
detector material.

 Knowing parameters, characterizing the inhomogeneity of materials, one can
calculate the transport of slow neutrons in the medium and the neutron albedo
from this medium.

 Figures. 6 and 7 demonstrate examples of homogeneous materials that do not
show exhibit deviation from the inverse velocity law in their total cross
sections.

 The total cross section curves (Fig. 6) for Si-wafers and fluoropolymers at
room temperature (liquid state) and low temperatures (90 K, solid state) do
not show any sign of inhomogeneities.
 The measurements of the angular distributions of neutron scattering for these
materials have not demonstrated any broadening of the collimated UCN beam.

 The total cross section of silicon (Fig. 6) is in good agreement with the
cross section extrapolated from the energy range of cold neutrons
\cite{Freu,Mug} according to $1/v$ law and can be fitted by the dependence
\begin{equation}
\sigma=\sigma_{0}+a\cdot\lambda^{'},
\end{equation}
where $\lambda^{'}$ is the neutron wavelength in the substance (corrected for the Fermi potential of the sample
material), with parameters $\sigma_{0}=(3.26\pm 2.3$) b, $a=(0.122\pm 0.005)$ b/\AA.
 This cross section is the sum of the energy-independent elastic coherent
scattering and the cross section of neutron capture and heating, which are proportional
to the neutron wavelength.

 The low temperature liquid fluoropolymer POM-310\footnote{The investigated
substances were produced by the Perm' branch of the Russian Scientific Center "Applied Chemistry"} is the
mixture of complex polyfluorooxymethylenes having the general formula
CF$_{3}$O(CF$_{2}$O)$_{n}$(CF$_{2}$CF$_{2}$O)$_{m}$(OCF$_{2}$CF$_{2}$O)$_{l}$ CF$_{3}$ with $n:m:l=65.8:3.1:0.2$
and a molecular weight 4883.
 The measured total neutron cross section for this polymer yields the following
parameters for Eq. (14): $\sigma_{0}=(1.53\pm 2.2$) b/atom, $a=(8.8\pm 0.5)\times 10^{-2}$ b/(atom \AA) at 300
K, and $\sigma_{0}=(-3\pm 3$) b/atom, $a=(4.2\pm 0.5)\times 10^{-2}$ b/(atom \AA) at 90 K.

 Fig. 7 shows the total cross section for polyethylene film.
 No visible deviation from $1/v$ law is observed at the degree of precision of
the test measurements, but our tentative measurement of the differential cross section reveals certain elastic
scattering from density inhomogeneities even for such a material with small strength of coherent scattering
(U$\approx$-9 neV).
 The fit of the total cross section data in accordance with Eq. (14)
provides: $\sigma_{0}=(96\pm 18$) b, a=$(2.07\pm 0.04$) b/(atom \AA).
 From the measurements, it follows that with account of the neutron capture,
extrapolated to the thermal point UCN upscattering contribution is
$\sigma_{ups,extr}=(2.07\pm 0.4)\times 1.8-(0.334\times 2/3)=
(3.5\pm 0.08)$b/atom of the (CH$_{2}$)-complex.

 Fig. 8 presents the example of the measured differential cross section for
copper.

 The parameters of sample inhomogeneity appeared to differ from the parameters
obtained from the total cross section:
Cu (Gaussian): $G_{0}=(18\pm 4) neV^{2}, \rho_{0}=(420\pm 45)$\, \AA.
 Analysis of more sophisticated models describing density fluctuations in
inhomogeneous media calls for a more comprehensive investigation, which is
beyond the scope of this test study.

\section{Acqnowledgments}
The spectrometer test and measurements were performed in the course
of experiments 3-14-110, 3-14-132, and 3-14-152 at the PF2-test channel of the
UCN turbine source of the High Flux Reactor at the Institut Laue-Langevin
(Grenoble, France). We are grateful to the ILL reactor personal.


\begin{thebibliography}{40}
\bibitem{ucn}
 F. L. Shapiro, In: {\it Proceedings of the International Conference
on Nuclear Structure with Neutrons, Budapest, 1972} edited by J. Ero
and J. Szucs (Plenum, New York, 1972), p.259;
 A. Steyerl, in {\it Neutron Physics, Springer Tracts in Modern Physics,
{\bf 80}}, (Springer, Berlin, Heidelberg, New York, 1977), p. 57;
 R. Golub and J. M. Pendlebury, Rep. Progr. Phys., {\bf 42} (1979) 439;
 V. K. Ignatovich, {\it Fizika ultrakholodnykh neitronov},
(Nauka, Moscow,1986, in Russian);
 {\it The Physics of Ultracold Neutrons}, (Clarendon, Oxford, 1990);
 R. Golub, D. J. Richardson and S. Lamoreaux, {\it Ultracold Neutrons}
(Adam Hilger, Bristol, 1991);
 J. M. Pendlebury, Ann. Revs. Nucl. Part. Sci., {\bf 43} (1993) 687.

\bibitem{ILL}
{\it Proceedings of the International Conference on Fundamental Physics
with slow Neutrons, Grenoble, 1998}, Nucl. Instr. Meth., {\bf A440} (2000)

{\it Proceedings of the International Conference on Precision Measurements
with slow Neutrons, Gaithersburg, USA, 2004}, Journ. of Res. NIST, {\bf 110},
(2005) No. 3 and 4.

{\it Proceedings of the International Workshop on Particle Physics with slow
Neutrons, ILL, Grenoble, France, 2008}, Nucl. Instr. Meth., {\bf A 611} (2009)
Iss. 2-3.

\bibitem{Gol}
R. Golub, Revs. Mod. Phys. {\bf 68} (1996) 329.

\bibitem{Mich}
A. Michaudon, Los Alamos Preprint LA-13197-MS, Los Alamos, 1997.

\bibitem{turb}
A. Steyerl, H. Nagel, F.-X. Schreiber, K.-A. Steinhauser, R. Gahler,
W. Gl\"aser, P. Ageron, J.-M. Astruc, N. Drexel, R. Gervais, W. Mampe,
Phys. Lett., {\bf A116} (1986) 347.

\bibitem{LA}
P. E. Hill, J. M. Anaya, T. J. Bowles, G. L. Greene, G. Hogan, S. Lamoreaux,
L. Marek, R. Mortenson, C. L. Morris, A. Saunders, S. J. Seestrom, W. Teasdale,
S. Hoedl, C.-Y. Liu, D. A. Smith, A. Young, B. W. Fillipone, J. Hua, T. Ito,
E. Pasyuk, P. Geltenbort, A. Garcia, B. Fujikawa, S. Baessler, A. Serebrov,
Nucl. Instr. Meth., {\bf A440} (2000) 674.

A. Saunders, J. M. Anaya, T. J. Bowles, B. W. Fillipone, P. Geltenbort, P. E. Hill, M.Hino, S. Hoedl,  G. Hogan, T. Ito,
K.W. Jones, T. Kawai, K. Kirch, S. Lamoreaux, C.-Y. Liu, M. Makela, L.J. Marek, J.W. Martin, C. L. Morris,
R.N. Mortenson, A. Pichlmaier, S.J. Seestrom, A. Serebrov, D. Smith, W. Teasdale, B. Tipton, R.B. Vogelaar,
A.R. Young, J. Yuan, Phys. Lett., {\bf B593} (2004) 55.

\bibitem{FRM}
U. Trinks, F. J. Hartmann, S. Paul, W. Schott, Nucl. Instr. Meth., {\bf 440}
(2000) 666.

\bibitem{PSI}
Workshop on PSI UCN Source, 17-18 Dec. 2001, PSI, Villigen, Switzerland;
http://ucn.web.psi.ch/techrev\_ucn/htm

\bibitem{D2}

I.S. Altarev, Yu.V. Borisov, A.B. Brandin et al., Phys. Lett., {\bf 80A} (1980) 413.

R. Golub, K. B\"oning K, Z. Phys. {\bf B51} (1983) 95;

Z.Ch. Yu , S.S. Malik, R. Golub, Z. Phys. {\bf B62} (1986) 137;


I.S. Altarev, V.A. Mityukhlyaev, A.P.  Serebrov, A.A. Zakharov, Journ. Neutr. Res., {\bf 1} (1993) 71.




A.P. Serebrov, Nucl. Instr. Meth. {\bf A440} (2000) 653;

A. Serebrov A., A. Mityukhlyaev, A. Zakharov et al. Nucl. Instr. Meth. {\bf A440} (2000) 658.

\bibitem{horspe}
M.G.D. van der Grinten, J.M. Pendlebury, D. Shiers, Nucl. Instr. Meth.
{\bf A423} (1999) 421;

F. Atchison, B. Blau, M. Daum et al., Phys. Lett. {\bf B642} (2006) 24;

F. Atchison, B. Blau, M. Daum, et. al., Nucl. Instr. Meth. {\bf B260} (2006)
647;

I. Altarev, M. Daum, A. Frei et al., Eur. Phys. Journ. {\bf A37} (2008) 9.

\bibitem{hor-my}
M.I. Novopoltsev, Yu.N. Pokotilovski, JINR Commun. Ð3-81-828, 1981, Dubna;

M.I. Novopoltsev, Yu.N. Pokotilovski, I.G. Shelkova,  Nucl. Instr. Meth,
{\bf A264} (1988) 518;

M.I. Novopoltsev, Yu.N. Pokotilovski, Prib. Tekhn. Exper., no. 5 (2010) 19
[Instrum. Exp. Tekhn. {\bf 53} (2010) 635]; arXive: 1008.1419.

\bibitem{Stespe}
A. Steyerl, Phys. Lett. {\bf 29B} (1969) 33;

A. Steyerl and H. Vonach, Z. Phys. {\bf 250} (1972) 166;

A Steyerl, Nucl. Instr. Meth. {\bf 101} (1972), 295.

\bibitem{Perspe}
A.V. Antonov, A.V. Isakov, S.P Kuznetsov et. al., Fiz. Tverd. Tela {\bf 26}
(1984) 1585.

\bibitem{CH2}
Yu.Yu. Kosvintsev, Yu.A. Kushnir, V.I. Morozov, ZhETF {\bf  77} (1979) 1277
[JETP {\bf 50} (1979) 642];

A. Steyerl, S.S. Malik, P. Geltenbort et al., J. Phys., III, France, {\bf  7}
(1997) 1941;

T. Bestle, P. Geltenbort, H. Just, Phys. Lett. {\bf A244} (1998) 217;

L.N. Bondarenko, P. Geltenbort, E. Korobkina et. al, Pis'ma ZhETF {\bf 68}
(1998) 663;

P. Geltenbort, V.V. Nesvizhevsky, D.G. Kartashov et.al., Pis'ma ZhETF {\bf 70}
(1999) 175;

V.V. Nesvizhevsky, A.V. Strelkov, P. Geltenbort, P. Iaydjiev, Eur. Phys. Journ.
{\bf AP6} (1999) 151;

E.V. Lychagin, A.Y. Muzychka, V.V. Nesvizhevsky et al., Phys. Lett. {\bf B479}
(2000) 353;

L.N. Bondarenko, P. Geltenbort, E. Korobkina et. al., Jad. Fiz., {\bf 65} (2002)
13.

E.V. Lychagin, D.G. Kartashov, A.Y. Muzychka, et al, Jad. Fiz. {\bf 65} (2002)
2052.

\bibitem{Rich}
D.J. Richardson, J.M. Pendlebury, P. Iaydjiev, Nucl. Instr. Meth. {\bf A308} (1991)
568.

\bibitem{PFPE}
J.C. Bates, Phys. Lett. {\bf 88A} (1882) 427; Nucl. Instr. Meth. {\bf A216} (1983) 535;

P. Ageron, W. Mampe, J.C. Bates, and J.M. Pendlebury, Nucl. Instr. Meth.
{\bf A249} (1986) 261;

W. Mampe, P. Ageron, J.C. Bates et al., Nucl. Instr. Meth. {\bf A284} (1989)
111; Phys. Rev. Lett. {\bf 63} (1989) 593;

Yu.N. Pokotilovski, Nucl. Instr. Meth. {\bf A425} (1999) 320;

S. Arzumanov, L. Bondarenko, S. Chernyavsky et al., Phys. Lett. {\bf B483}
(2000) 15;

A. Pichlmaier, J. Butterworth, P. Geltenbort et al., Nucl. Instr. Meth.
{\bf A440} (2000) 517;
A. Pichlmaier, Dissertation. (TU M\"unchen, 1999);

Yu.N. Pokotilovski, ZhETF {\bf 123} (2003) 203 [JETP {\bf 96} (2003) 172];

A. Serebrov, V. Varlamov, A. Kharitonov et al., Phys. Lett. {\bf B605} (2005)
72, Phys. Rev. {\bf C78} (2008) 035505;

Yu.N. Pokotilovski, I. Natkaniec, K. Holderna-Natkaniec, Physica {\bf B403}
(2008) 1942.

\bibitem{LiF}
C.A. Baker, K. Green, M.G.D. van der Grinten et al., Nucl. Instr. Meth.
{\bf A487} (2002) 511;

C.A. Baker, S.N. Balashov, K. Green et al., Nucl. Instr. Meth. {\bf A501}
(2003) 517;

T. Kitagaki, K. Sakai, M. Hino et al., Nucl. Instr. Meth. {\bf A529} (2004)
425;

M. Lasakov, A. Serebrov, A. Khusainov et al., Journ. Res. Natl. Inst. Technol.
{\bf 110} (2005) 289.

\bibitem{Leng}
M. Lengsfeld and A. Steyerl., Z.Phys. {\bf B27} (1977) 117;

A. Lermer, A. Steyerl, Phys. Stat. Sol. {\bf A33} (1976) 531.

\bibitem{inhom}
A. Steyerl, {\it Proc of the 2nd Intern. School on Neutron Physics}, Alushta,
1974, JINR, Dubna, D3-7991, 42;
{\it Springer, "Very Low Energy Neutrons", Springer Tracts in Modern Physics},
Berlin, Heidelberg, N-Y.: Springer, {\bf 80} (1977) 57;

A.V. Stepanov, Fiz. Elem. Chast. i Atom. Jad. {\bf 7} (1976) 989.

\bibitem{Perfor}
V.G. Grinev, A.I. Isakov, S.P. Kuznetsov et. al., Journ. Mosc. Phys. Soc.,
{\bf 2} (1992) 243.

\bibitem{Gran}
G.J. Cuello, J.R. Santisteban, R.E. Mayor et al., Nucl. Instr. Meth., {\bf A357}
(1995) 519.

\bibitem{Freu}
A.K. Freund, Nucl. Instr. Meth. {\bf 213} (1983) 495.

\bibitem{Mug}
S.F. Mughabghab, M. Divadeenam, N.F. Nolte, Cross Sections {\bf Vol.1}
Part A. Academic Prees INC. 1981;
S.F. Mughabghab, Cross Sections {\bf Vol.1} Part B. Academic Press INC. 1984.

\end{thebibliography}
\end{document}